\documentclass[twoside]{LCWS11}
\usepackage[latin1]{inputenc}
\usepackage[dvips]{graphicx,epsfig,color}
\usepackage{wrapfig,rotating}
\usepackage{amssymb,amsmath,array}
\usepackage{cite}
\usepackage{slashed}
\usepackage{feynarts}

\newcommand{\bea}{\begin{eqnarray}}
\newcommand{\eea}{\end{eqnarray}}
\newcommand{\bean}{\begin{eqnarray*}}
\newcommand{\eean}{\end{eqnarray*}}
\newcommand{\nn}{\nonumber \\}

\def\gb #1{ \left\langle #1 \right]}

\def\Label#1{\label{#1} \smash{\hbox to0pt{\raise1ex\hbox{\tiny[#1]}\hss}}}

\newcommand{\be}{\begin{equation}}
\newcommand{\ee}{\end{equation}}
\newcommand{\bd}{\begin{displaymath}}
\newcommand{\ed}{\end{displaymath}}

\begin{document}
\title{
%%%%   Paper title goes here  %%%%%%%%%%%%%%
 Massive particles and unitarity cuts} %% 
%***********************************************************************
% AUTHORS INFORMATION AREA
%***********************************************************************
\author{Ruth Britto$^1$ and Edoardo Mirabella$^2$
% Optional short acknowledgment: remove next line if non-needed
%\thanks{This is an optional funding source acknowledgment.}
% DO NOT MODIFY THE FOLLOWING '\vspace' ARGUMENT
\vspace{.3cm}\\
% Addresses and institutions (remove "1- " in case of a single institution)
1- Institut de Physique Th\'eorique, CEA-Saclay, \\
F-91191, Gif-sur-Yvette cedex, France 
\vspace{.1cm}\\
2- Max-Planck-Institut f\"ur Physik, \\
F\"ohringer Ring 6, D-80805 M\"unchen, Germany\\
}
%%***********************************************************************
% END OF AUTHORS INFORMATION AREA
%***********************************************************************

\maketitle
 
\begin{abstract}
We present an extension of   the spinor integration formalism of one loop amplitudes from 
the double-cut to the single-cut case. This technique can be applied for the computation 
of the tadpole coefficients. Moreover we describe an  off-shell continuation of 
 one loop amplitudes  that allows a finite evaluation of the unitarity cuts
in the channel of a single massive external fermion.
\end{abstract}

%%***********************************************************************
% INTRODUCTION
%***********************************************************************

\section{Introduction}
Recent years have seen rapid progress in computing one-loop amplitudes, due largely to the use of 
generalized unitarity methods~\cite{Cutkosky:1960sp,vanNeerven:1985xr,Bern:1994zx,Bern:1994cg,Bern:1997sc,Bern:2004ky,Britto:2004nc,Mastrolia:2006ki,Forde:2007mi,Ossola:2006us,Ellis:2007br,Kilgore:2007qr,Giele:2008ve,Berger:2008sj,Badger:2008cm,Ellis:2008ir}. These methods   enable the computation of loop amplitudes in terms of tree-level amplitudes, 
which are (relatively) easy to generate either numerically or analytically.  They  rely   
on the knowledge of the expansion of any amplitude in terms of a  set of master integrals,  with coefficients that are  rational functions of the kinematic invariants. Indeed  they 
operate by matching the  generalized cuts of the loop  amplitude and the master integrals.

Unitarity methods work most elegantly  when all internal particles are massless.  If massive particles 
are involved, there are new master integrals whose cuts are more difficult to solve.
The  additional master integrals are the tadpoles $A_0(m)$, and the ``on-shell bubbles'' $B_0(m^2;m,0)$ and 
$B_0(0;m,m).$\footnote{These bubbles are called on-shell because the momentum involved in the cut 
channel is the on-shell momentum of an external particle.}  Several methods, either numerical or analytical, 
aim to   compute   these coefficients~\cite{Ossola:2006us, Mastrolia:2010nb,Ellis:2011cr,Bern:1995db, Badger:2008za,Britto:2009wz,Britto:2010um}.  

% SE METTO ANCHE IL SINGE CUT
In the following  we will be focus on  the tadpole integral and on the bubble $B_0(m^2;m,0)$,  presenting
methods for their analytical computations.   More detailed  discussions can be found 
elsewhere~\cite{Britto:2010um, Britto:2011cr}.

%%***********************************************************************
%  SINGLE CUT
%***********************************************************************

\section{Single cut integration}
\label{Sec:SingleCut}
The tadpole coefficient can be computed by cutting just one propagator. In this section we describe an extension of the  formalism for explicit evaluation of double cuts
 \cite{Britto:2005ha,Britto:2006sj,Anastasiou:2006jv,Britto:2006fc,Anastasiou:2006gt,Britto:2007tt,Britto:2008vq,Feng:2008ju,Mastrolia:2009dr} to the single cut case~\cite{Britto:2010um}.
 The loop momentum is parametrized in terms of spinor variables and the $(D-2)$-dimensional  integral is performed algebraically by the Cauchy residue theorem.
 We find  that  the full single cut integral will typically diverge. Moreover,
 while  the evaluation of the double cut  involves the computation of  residues at poles,   in the evaluation of the single cut the contour integral part of the formula 
 dominates the tadpole contribution,  so the  residues are not needed.   Finally in the single cut case it is convenient to work at the integrand level.
 
The starting point is the one-loop integrand, 
\be
I = \frac{N(\ell)}{D_0 D_1 \cdots D_k},
\ee
where $N(\ell)$ is a polynomial in the loop momentum $\ell$, $ D_i=(\ell-K_i)^2 - m_i^2$, and $K_0 =0$. 
 The 4-dimensional single-cut operator for the propagator  $D_0$  acts on the integrand as
\bea
\Delta_{D_0}[I] \equiv 
\int d^4\ell~  \delta^{(+)}\left (\ell^2-m_0^2 \right ) \; I_0, \qquad  I_0=\frac{N(\ell)}{D_1 \cdots D_k}. 
\label{operator-def}
\eea
The single cut is applied to the  {\it integrand}, because of the presence of   non-vanishing contributions from the  spurious terms.  Working 
with the  integrand allows  us to identify the particular propagator being cut.  The single cut~(\ref{operator-def}) can be written as follows:
\bea 
\Delta_{D_0}[I] 
= 
\int_0^\infty  \frac{dt}{4} \int \left(i dz \wedge d\bar{z} \right)
\frac{K^2 t^2 (1+z\bar{z}) }
{ \sqrt{ t^2 (1+z\bar{z})^2+u }} I_0, 
\label{cut-var}
\eea
where $z$, $\bar z$  and $t$ are defined according to
\be
\ell^\mu =  t \left( p^\mu + z \bar{z} q^\mu + \frac{z}{2}\gb{q|\gamma^\mu|p}
- \frac{\bar z}{2}\gb{p|\gamma^\mu|q} \right) +   \frac{u}{2}
\frac{1}
{\sqrt{t^2 (1+z\bar{z})^2+u } +  t(1+z\bar{z}) } K^\mu \;  , 
\ee
and $u \equiv 4 m_0^2 / K^2$.  The momenta $p$ and $q$   are light-like and  
arbitrary,  while $K = p+q$.  They are 
chosen  such that $K^2 \gg m_0^2$ since it is convenient to work in the 
 $u \to 0$ limit.  The integration over $z$ and $\bar{z}$ will be performed by the Generalized Cauchy Formula as 
described in \cite{Mastrolia:2009dr}.  For the integrand $F(z,\bar{z})$ we construct a primitive $G(z,\bar{z})$ with respect to, say, $\bar{z}$. Then
\bea
\int_D   F(z,\bar{z}) ~d\bar{z} \wedge dz
&=& \oint_{\partial D} dz~ G(z,\bar{z}) 
- 2 \pi i \sum_{{\rm poles}~ z_j } {\rm Res} \{G(z,\bar{z}),z_j \}  \nn
&=& \int_{0}^{2 \pi} d\alpha~ \Lambda e^{i \alpha}G(\Lambda e^{i \alpha}, \Lambda e^{-i\alpha}) 
- 2 \pi i \sum_{{\rm poles}~ z_j } {\rm Res} \{G(z,\bar{z}),z_j \} 
\ ,
\label{gcf}
\eea
where  $D$ is a disk of radius $\Lambda$ encompassing the poles of $G(z,\bar{z})$:
\bean
z=re^{i\alpha}; \qquad D=\{(r,\alpha) ~|~ 0\leq r \leq \Lambda, ~0 \leq \alpha < 2 \pi \}.
\eean
The information we need is enclosed at the integrand level, thus the integration over $t$ will not be carried out.
 It turns out that in the limit $\Lambda \to \infty$, we can ignore the residues and restrict our attention to the closed line integral term.  In this limit, the
 leading behavior of the primitives of the master  integrands is given by
 \be
 I=\frac{1}{D_0}  \; \Longrightarrow  \; \Lambda e^{i \alpha}G  \stackrel{\Lambda \to \infty}{=}  \Lambda^2 t \; , \qquad  \qquad
 I= \frac{1}{D_0 \cdots D_n}  \; \Longrightarrow  \; \Lambda e^{i \alpha}G  \stackrel{\Lambda \to \infty}{=}  \frac{\log \Lambda^2}{\Lambda^{n-2}},  
  \ee
  with $n=2,3,4$. The higher-point integrands are suppressed by powers of $\Lambda$.  Moreover tadpole primitives are purely 
  rational, while the others are purely logarithmic.  Therefore, in an algorithm targeting tadpole coefficients, 
  we will select  terms of the single cut with specific dependence on $\Lambda^2$.

\subsection*{Computation of tadpole coefficients}
In this section we show how the single cut allows the computation of tadpole coefficients.
The idea is to expand the integrand in The Ossola-Papadopoulos-Pittau  (OPP) 
decomposition~\cite{Ossola:2006us}, in order to easily recognize the spurious contributions.
The coefficients of both spurious and physical terms are treated as unknowns, and 
the physical terms except for the tadpole are dropped.  Single cuts of all remaining terms are 
then evaluated. Thanks to the OPP  expansion, the single-cut 
equation becomes a system of separate equations, which are  the coefficients of independent tensors.

As a simple example, we compute the coefficient $a(0)$ of the tadpole integral $A_0(m_0^2)$ of the integrand
\be
I = { 2 \ell \cdot R \over D_0 D_1}.
\ee
We assume its Gram determinant  is nonvanishing, i.e. $K_1^2 \neq 0$, and that  the masses 
are non-degenerate.\footnote{When this is not the case, further modifications are in order~\cite{Britto:2009wz}.}
The OPP decomposition of $I$  is  given by
\begin{equation}
I = \frac{a(0)}{D_0} +  \tilde b_{11}(01) \, \frac{2 \ell \cdot  \ell_7}{ D_0D_1} +   \tilde b_{21}(01) \, \frac{2 \ell \cdot  \ell_8}{ D_0D_1} +
 \tilde b_{0}(01) \, \frac{2 \ell \cdot  n}{ D_0D_1} + \cdots.
 \label{eq:I21start}
\end{equation}
Terms whose single cut contains no  $\Lambda^2t$ contribution are included in ``$\cdots$''  and are neglected. 
The momenta  $n$, $\ell_7$ and $\ell_8$ are defined~\cite{Ossola:2006us} to satisfy the conditions
\begin{displaymath}
K_1 \cdot n = K_1 \cdot \ell_7 =K_1 \cdot \ell_8 = 0,~~~~~ n^2 = \ell_7 \cdot \ell_8 = -K_1^2,~~~~~ \ell_7^2 = \ell_8^2 =0.
\end{displaymath}
 Applying the single cut operator $\Delta_{D_0}$ and selecting the $\Lambda^2 t$ terms only, we get 
\begin{eqnarray}
0 &=&\Bigg [  - \left (a(0)+\alpha_1 \right ) K_1^\mu  +  \left  ( \tilde b_{11}(01)-\alpha_3 \right )  \ell^\mu_7 \nonumber  \\ 
&+&  \left ( \tilde b_{21}(01) -\alpha_4  \right )  \ell^\mu_8 + \left (  \tilde b_{0}(01) -\alpha_2 \right )  n^\mu \Bigg ] \; \frac{ q_\mu}{ K_1 \cdot  q } \; 
\Delta_{D_0}\left [\frac{1}{D_0} \right  ]. 
\label{eq:I21main}
\end{eqnarray}
The coefficients $\alpha_{i=1,\cdots,4}$ are the coordinates of  $R$ in  the basis $\{ K_1, n, \ell_7, \ell_8 \}$. They read as follows:
\begin{equation}
\alpha_1 = \frac{R \cdot K_1}{K_1^2}, ~~~\alpha_2 = - \frac{R \cdot n}{K_1^2},~~~\alpha_3 =  - \frac{R \cdot \ell_8}{K_1^2}, ~~~\alpha_4= - \frac{R \cdot \ell_7}{K_1^2}.
\end{equation}

Since $q$ is arbitrary, the expression inside the square brackets has to vanish.  Therefore each of the factors  multiplying the basis vectors vanishes separately, giving four equations.  
The tadpole coefficient is obtained from the  first of these equations,
\begin{equation}
a(0)+\alpha_1 = 0 ~~\Longrightarrow~~ a(0) = - \frac{R \cdot K_1}{K_1^2}.
\end{equation}
This result has been checked against the one obtained from the Passarino-Veltman decomposition.

%%***********************************************************************
%  EXTERNAL LEG CORRECTIONS
%*****************************************************************

\section{External leg corrections in unitarity methods}
%  SE C' E' IL SINGLE CUT!!!!!
%The unitarity cut of the amplitude of a  process with massive external particles should give information about the coefficients of the 
%on-shell Green's function. Unfortunately the external leg correction diagrams are singular, because the propagator opposite 
%the external leg carries the same momentum and is therefore also on shell.     
%\medskip
%
The unitarity cut of the amplitude of a  process with massive external 
particles should give information about the coefficients of the  on-shell Green's function. Unfortunately the external 
leg correction diagrams are singular, because the propagator opposite  the external leg carries the same 
momentum and is therefore also on shell.  This problem was addressed 
 in  the context of a numerical algorithm~ \cite{Ellis:2008ir,Ellis:2011cr}.  The solution proposed 
was to omit the problematic  contributions, taking care with the associated breaking of 
gauge invariance. 

Here, we present a different solution~\cite{Britto:2011cr}.
 In the spirit of the unitarity method, we want to keep the 
ingredients of the cut as complete amplitudes, without discarding any  contributions.  We must 
therefore also include the corresponding counterterms. 
The examples given here involve 4-dimensional cuts, although the formalism is equally valid in $D$ dimensions.

\noindent
We regularize the   unitarity cuts by  continuing  the amplitude off-shell.  In particular we perform a double momentum shift modifying the 
massive external momentum $k$ and one other external momentum $r$. If both momenta are outgoing, the shift is
\bea
k \to \hat k = k +   \xi \bar k, \qquad 
r \to \hat r = r -  \xi \bar k,
\label{shift}
\eea
where the momentum $\bar k$  is such that
$r \cdot \bar k =  \bar k^2=0$ and  $2(k \cdot \bar k)  \neq0$.  
The momenta $k$  and $\hat r$ are on shell, while $\hat{k}$ is off shell. 
 With this shift, the the full amplitude diverges  as $1 / \xi$. 
 The cuts are calculated in terms of tree amplitudes. For the double cut, one has simply the off shell  three-point 
 interaction $\mathcal{M}_R$ and  an $(n+1)$-point  on shell tree amplitude $\mathcal{M}_L$,  
 the latter  depending on the parameter $\xi$.  In the case of the single cut we need a single $(n+2)$-point $\xi$-dependent
  tree amplitude, $\mathcal{M}_T$, which is continued of shell.  We 
 expand $\mathcal{M}_L$ and $\mathcal{M}_T$  up to  first order in $\xi$.

In the on-shell scheme, the external leg correction diagrams are exactly cancelled 
by the corresponding counterterms. The  counterterm diagrams  have to be  constructed 
from the renormalization constants and the various tree-level off-shell currents.  
After the  expansion  in $\xi$,  the divergent part is guaranteed to  
cancel the cut loop diagram and  the on-shell limit is reached by setting  $\xi =0$. 
Therefore at every stage of the procedure 
we systematically neglect terms of $\mathcal{O}(\xi)$.

The off-shell currents are gauge-dependent.  In the sum of all parts, gauge invariance is restored by construction:  coefficients of master integrals are gauge invariant, and 
we have only added zero in the form of the external leg correction plus its counterterm.

In the following we will  show  the cancellation of divergences in the cut 
between loop Feynman diagrams and the counterterms. 
\subsection*{Bubbles from double cut}
\label{SSSec:BubDC}
  
\begin{figure}[t]
\input{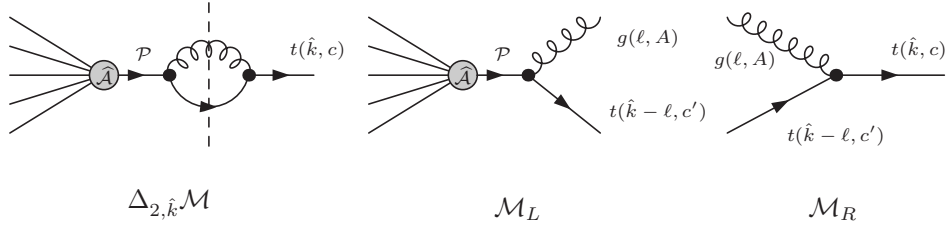}
\caption{Double cut of the external leg correction diagram, and the left and right tree-level amplitudes. 
The momentum of the external massive fermion, $k$, is outgoing.  
The cut momenta are $\ell$ and $k-\ell$.  Color information is indicated by $c$ and $c'$.  The massive propagator giving the on-shell divergence is denoted by $\mathcal{P}$.}
\label{Fig:DoubleCut}
\end{figure} 
  
We consider the double cut of the external leg correction diagram for a massive fermion, 
as shown in  Figure~\ref{Fig:DoubleCut}.  The tree-level amplitudes $\mathcal{M}_L$ and $\mathcal{M}_R$
depicted in Figure~\ref{Fig:DoubleCut} read as follows:
\begin{eqnarray}
\mathcal{M}_L &=&  \frac{ g \, T^A_{c' c''}  }{(k+  \xi \bar k)^2-m^2 } \; \left ( \bar u_{k+   \xi \bar k-\ell} \,  \slashed{\varepsilon}^{\ast}_{\ell} \,
(m+\slashed{k}+   \xi \slashed{\bar k} ) \, \widehat{\cal A}_{c'' c_{\rm ext}}  \right ), \nonumber \\
\mathcal{M}_R &=& -i\,g\,T^A_{cc'}   \left ( \bar u_{k} \,  \slashed{\varepsilon}_{\ell} \, u_{k 
+  \xi \bar k-\ell} \right ),
\label{Mleftright}
\end{eqnarray}
where  $\widehat{\cal A}_{c'' c_{\rm ext}}$ is  the remaining parts of the diagram, including the color-flow information. It is worth noticing
that the external massive spinor $u_k$ is {\em not} being shifted.\footnote{As explained in~\cite{Britto:2011cr}, the  shift of $u_k$ is possible but  unnecessary.}
In the Feynman gauge,    the  double cut including the sum over the polarization states is
\bean
&&- \frac{2 i g^2 C_F}{ \xi   \gamma} 
{1 \over (2 \pi)^4}  \int d^4 \ell \delta(\ell^2) \delta\left ( ( \ell - \hat k) - m^2 \right ) \nn
&& \qquad \Big [
(2 m^2 - \xi   \gamma)  \left ( \bar u_k \, \widehat{\mathcal{A}}_{cc_{\rm ext}}  \right ) +
 \left ( \bar u_k \, \slashed{\ell} \, (m+\slashed k +   \xi \slashed{\bar k})\,   \widehat{\mathcal{A}}_{cc_{\rm ext}}  \right ) +
 m   \xi \left ( \bar u_k \, \slashed{\bar k} \, \widehat{\mathcal{A}}_{cc_{\rm ext}}  \right ) 
\Big ].
\eean
The Dynkin index of the fundamental representation of $SU(N)$ is denoted by $C_F$, while  $\gamma \equiv 2 k \cdot \bar k$.
 
 Computing  the double cut and performing the $\xi$-expansion,  we get the bubble part of the diagram:
\begin{eqnarray}
\mathcal{M}_B &=& \frac{g^2 C_F}{16 \pi^2   \xi \gamma} \Bigg \{    
4 m^2   \xi \gamma \left (  \bar u_k \,{\mathcal{A}}_{cc_{\rm ext}}  \right )  B'_0(m^2;m,0) \nonumber 
\\
&+&   \bigg [ 
4 m^2 \left (  \bar u_k \,{\mathcal{A}}_{cc_{\rm ext}}  \right )  +
4m^2   \xi  \left (  \bar u_k \,  {\mathcal{A}}'_{cc_{\rm ext}}  \right ) +
2m   \xi  \left (  \bar u_k \,\slashed{\bar k} \, {\mathcal{A}}_{cc_{\rm ext}}  \right )
\bigg ] B_0(m^2;m,0)
  \Bigg \}  \ .
 \label{Eq:DCexpanded}
\end{eqnarray}
 In obtaining eq.~(\ref{Eq:DCexpanded}) we have used the expansions 
\begin{eqnarray}
&& B_0(m^2+  \xi \gamma;m,0) =  B_0(m^2;m,0)+ \xi  \gamma  B'_0(m^2;m,0) \ ,  \nn
&& \widehat{\mathcal{A}}_{cc_{\rm ext}}  =  {\mathcal{A}}_{cc_{\rm ext}} +\xi {\mathcal{A}}'_{cc_{\rm ext}}  \ . \nonumber
\end{eqnarray}

When the amplitudes are written  in the spinor-helicity formalism the completeness relation for polarization vectors is that of
a light-like axial gauge rather than Feynman gauge. Therefore the  procedure described  above must be  modified. In particular 
the bubble part of the diagram, eq.~(\ref{Eq:DCexpanded}),
gets an extra $\mathcal{O}(\xi^0)$ contribution
\begin{equation}
\frac{g^2 C_F}{16 \pi^2 \gamma} \frac{\bar u_k \,  \slashed{\bar k} \, \slashed{k} \, \slashed{q}  (m+\slashed{k})   \, {\mathcal{A}}_{c c_{\rm ext}}   }{q \cdot k}  B_0(m^2;m,0),
\label{Eq:ExtraDouble}
\end{equation}
where $q$ is an arbitrary light-like ``reference'' momentum. This new contribution does not affect the divergent part of the bubble coefficient.

\begin{figure}[t]
\input{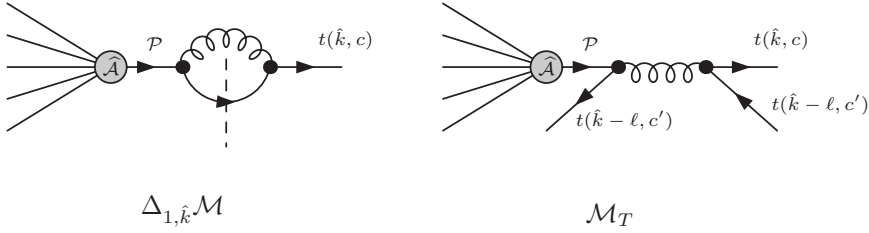}
\caption{Single cut of the external leg correction diagram.}
\label{Fig:SingleCut}
\end{figure}

\subsection*{Tadpole from single cut}
  
The single cut of the massive propagator of the external leg correction diagram is depicted in 
Figure~\ref{Fig:SingleCut}.  The tree-level amplitude $\mathcal{M}_{T}$ is 
\begin{eqnarray}
\mathcal{M}_{T} &=&-\frac{ g^2 C_F}{ \ell^2   \xi \gamma} 
\left (\bar u_{k}\, \gamma^\mu \, u_{k +   \xi \bar k -\ell} \right) 
\left( \bar u_{k +   \xi \bar k - \ell} \, \gamma_\mu \, (m+ \slashed{k} 
+   \xi \slashed{\bar k} 
 ) \, \widehat{\mathcal{A}}_{c'c_{\rm ext}} \right ).
\end{eqnarray}
The single cut reads as follows
\bean
&& -\frac{2\, g^2 C_F}{   \xi \gamma}
{i \over (2 \pi)^4} \int d^4 \ell \ \delta(\ell^2) \nn
&& \qquad \left[\frac{
(2 m^2 -  \xi \gamma)  \left ( \bar u_k \, \mathcal{A}_{cc_{\rm ext}}  \right ) +
 \left ( \bar u_k \, \slashed{\ell} \, (m+\slashed k +   \xi \slashed{\bar k})\,   \mathcal{A}_{cc_{\rm ext}}  \right ) +
 m   \xi \left ( \bar u_k \, \slashed{\bar k} \, \widehat{\mathcal{A}}_{cc_{\rm ext}}  \right )}{\ell^2} 
\right ] \ ,
\eean
and can be computed using the method described in section~\ref{Sec:SingleCut}.
Expanding  around $  \xi = 0$ the  tadpole portion  of the external leg correction,  we get
\be
\mathcal{M}_A =  \frac{g^2 C_F}{16 \pi^2}  \bigg [
\left ( \frac{2}{  \xi\gamma} - \frac{1}{m^2} \right )   \left (  \bar u_k \, {\mathcal{A}}_{cc_{\rm ext}}   \right) + 
\frac{1}{\gamma m}   \left (  \bar u_k \,  \slashed{\bar k} \, {\mathcal{A}}_{cc_{\rm ext}}   \right) 
+\frac{2}{\gamma} \left (  \bar u_k \, {\mathcal{A}}'_{cc_{\rm ext}}   \right) 
 \bigg ]\;  A_0(m) .
 \label{Eq:SCexpanded}
\ee

\subsection*{Cancellation against the counterterm}
\label{SSec:Cancellation}
\begin{figure}[t]
\input{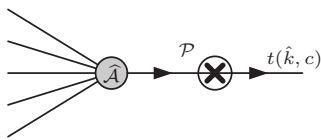}
\caption{Counterterm diagram.}
\label{Fig:CTdiag}
\end{figure}

The external leg counterterm, $\mathcal{M}^{\rm ct}$, depicted in Figure~\ref{Fig:CTdiag}  is
\bea
\mathcal{M}^{\rm ct} = -\frac{1}{  \xi\gamma} \Big( 
\bar u_k \,  \left (  \slashed{k}  \delta Z_{\psi} +  \xi \slashed{\bar k}  \delta Z_{\psi}-m\delta Z_\psi - m \delta Z_m\right ) \,
 (\slashed{k}+   \xi\slashed{\bar k}+ m)\, \widehat{\cal A}_{c c_{\rm ext}} 
\Big) .
\label{Eq:CT0}
\eea
In the on-shell scheme,  the  renormalization constants $\delta Z_m$ and $\delta Z_\psi$  read as follows:
\begin{eqnarray}
\delta Z_m &=&  -{g^2 C_F \over 16\pi^2} \left[ {A_0(m) \over m^2} + 2 B_0(m^2;m,0) \right], \nonumber \\
\delta Z_\psi &=& -{g^2 C_F \over 16\pi^2} \left[{A_0(m) \over m^2} - 4 m^2 B'_0(m^2;m,0) \right].
\end{eqnarray}
Expanding around $  \xi =0$ at $\mathcal{O}(  \xi^0)$, we get
\begin{eqnarray} 
\mathcal{M}^{\rm ct} &=& 
 -\frac{g^2 C_F}{16 \pi^2} \Bigg \{  
\bigg [ 
 \frac{2}{\xi   \gamma}A_0(m)  + \frac{4m^2}{\xi   \gamma}  B_0(m^2;m,0) \bigg  ]   \left (  \bar u_k \, {\mathcal{A}}_{cc_{\rm ext}}   \right)     \nonumber \pagebreak[1]  \\
&& \qquad \qquad 
 +
\bigg [ 
 \frac{2}{ \gamma}A_0(m)  + \frac{4m^2}{ \gamma}  B_0(m^2;m,0) \bigg  ]   \left (  \bar u_k \, {\mathcal{A}}'_{cc_{\rm ext}}   \right)     \nonumber \pagebreak[1]  \\
 && \qquad \qquad 
 +
\left [ \frac{1}{\gamma m} A_0(m) +\frac{2m}{\gamma} B_0(m^2;m,0) \right ]
\left (  \bar u_k \, \slashed{\bar k} \, {\mathcal{A}}_{cc_{\rm ext}}   \right ) 
 \nonumber \pagebreak[1]  \\
&& \qquad \qquad 
+ \bigg [ 
  \frac{1}{m^2}A_0(m) -   4m^2 B'_0(m^2;m,0)  \bigg  ]   \left (  \bar u_k \, {\mathcal{A}}_{cc_{\rm ext}}   \right)   
\Bigg\}. 
\label{Eq:CTexpanded0}
\end{eqnarray}
When the spinor-helicity  formalism  is used, the extra contribution~(\ref{Eq:ExtraDouble}) is 
accounted for by  adding the following term to eq.~(\ref{Eq:CT0}):
\begin{eqnarray}
\label{Eq:CTexpanded3}
\mathcal{M}^{k} &=& -\frac{1}{\xi \gamma}  \bar u_k  \bigg  [  (\slashed{k}+ \xi \slashed{\bar k} - m ) \, 
(\slashed{k} + \xi \slashed{\bar k} ) \slashed{q} \, \delta Z_{k}' \bigg ] (\slashed{k} + \xi \slashed{\bar 
k} + m)  \widehat{\mathcal{A}}_{c c_{\rm ext}},
\end{eqnarray}
where
\begin{eqnarray}
\delta Z_{k}'  &=& \frac{g^2 C_F}{16 \pi^2}  \frac{ B_0(m^2;m,0)}{q\cdot k}.
\end{eqnarray}

Once  %$\mathcal{M}_B$,  
eq.~(\ref{Eq:DCexpanded}), 
%$\mathcal{M}_A$, 
eq.~(\ref{Eq:SCexpanded}),  and 
%$\mathcal{M}^{\rm ct}$ 
eq.~(\ref{Eq:CTexpanded0})
are added together,  the terms proportional to  $1 /   \xi$  and the ones 
which depend on $\bar k$ cancel. The  sum is identically zero, since in the on-shell scheme the external leg correction diagram 
and   the external leg counterterm cancel  exactly.  The actual contribution to the tadpole and bubble coefficients 
comes from the other diagrams in the full amplitude. Since they are  finite in  $\xi$, no $\bar k$ dependence arises in the $\xi \to 0$ limit.

\section*{Acknowledgments}
This work was supported by the Research Executive Agency (REA) of the European Union under the Grant Agreement number PITN-GA-2010-264564 (LHCPhenoNet).
 This research was supported in part by the National Science Foundation under Grant No. NSF PHY05-51164;
we thank  the KITP  for its hospitality.   R.B. is supported by the Agence Nationale de la Recherche under grant ANR-09-CEXC-009-01. E.M. is supported by the European Research Council under Advanced Investigator Grant ERC-AdG-228301.

% ****************************************************************************
% BIBLIOGRAPHY AREA
% ****************************************************************************
\bibliographystyle{utphys}
\bibliography{references}
% ****************************************************************************
% END OF BIBLIOGRAPHY AREA
% ****************************************************************************

\end{document}